\documentclass[showpacs]{revtex4}

\begin{document}

\title{Involute, minimal, outer and increasingly trapped surfaces}

\author{Sean A. Hayward}
\affiliation{Center for Astrophysics, Shanghai Normal University, 100 Guilin 
Road, Shanghai 200234, China} 
\date{12th June 2009}

\begin{abstract}
Eight different refinements of trapped surfaces are proposed, of three basic 
types, each intended as potential stability conditions. Minimal trapped 
surfaces are strictly minimal with respect to the dual expansion vector. Outer 
trapped surfaces have positivity of a certain curvature, related to surface 
gravity. Increasingly (future, respectively past) trapped surfaces generate 
surfaces which are more trapped in a (future, respectively past) causal 
variation, with three types: in any such causal variation, along the expansion 
vector, and in some such causal variation. This suggests a definition of doubly 
outer trapped surface involving two independent curvatures. This in turn 
suggests a definition of involute trapped surface. Adding a weaker condition, 
the eight conditions form an interwoven hierarchy, with four independent 
relations which assume the null energy condition, and another holding in a 
special case of symmetric curvature.
\end{abstract}
\maketitle 

\section{Introduction} 
Trapped surfaces, as originally defined by Penrose \cite{Pen}, play an 
important role in gravitational physics, both for black holes and in cosmology. 
Such surfaces were crucial in the singularity theorems of Penrose and Hawking 
\cite{Pen,Haw,HP}. More recent years have seen the development of a local, 
dynamical theory of black holes in terms of marginal surfaces, a limit of 
trapped surfaces, including laws of black-hole dynamics involving physical 
quantities such as mass and surface gravity \cite{bhd,AK, Boo, Kri, 
GJ,bhd6,HCNVZ}. 

On the other hand, trapped surfaces as simply defined can have some very 
peculiar properties, such as the boundary of the region of trapped surfaces not 
being composed of marginal surfaces \cite{SK,Ben,BS}. As has been argued in a 
preliminary article \cite{ts1}, this suggests refining trapped surfaces in some 
natural way. Penrose already showed that the surface should be compact 
\cite{Pen}. The most obvious other refinement is to positive Gaussian 
curvature, which implies spherical topology if orientable, as will be assumed 
henceforth, but it does not suffice to resolve the above issue \cite{BS2}. It 
seems that a stability condition is required, which one might expect to be an 
inequality of one differential order higher than that of trapped itself, as for 
marginal surfaces \cite{New,KH,AMS,AMMS}. 

The preliminary article on spherical symmetry proposed seven such refinements, 
of essentially three types \cite{ts1}. In general cases, comparison of the 
definitions using the Einstein equation motivates an additional condition. The 
conditions are of some interest in themselves and turn out to be related: they 
form an interwoven hierarchy, with some direct relations and some relations 
which assume the Einstein equation, or more exactly, just the null energy 
condition (NEC). 

The article is organized as follows. Section II explains how any spatial 
surface in space-time is extremal in some normal direction, and defines minimal 
trapped surfaces. Section III defines outer trapped surfaces and shows that the 
condition is implied by minimal trapped. Section IV defines increasingly 
trapped surfaces and shows that the condition is implied by outer trapped. 
Section V defines doubly outer trapped surfaces, and shows that the condition 
implies both outer trapped and a stricter type of increasingly trapped. Section 
VI defines involute trapped surfaces, and shows that the condition implies both 
minimal trapped and doubly outer trapped. A laxer condition of involute trapped 
implies a laxer condition of increasingly trapped only in cases where outer 
trapped and doubly outer trapped are equivalent. The required geometrical 
quantities, such as expansion vector $H$, dual expansion vector $H_*$ and 
curvatures $K$ and $K_{(\pm)}$ related to surface gravity, are introduced as 
required, with concurrent comparisons to spherical symmetry, where the physical 
meanings are clearer. Section VII concludes with a hierarchy diagram, which one 
might well consult first. 

\vfill\eject
\section{Minimal trapped surfaces} 
The objects of study are spatial surfaces $S$ embedded in a given space-time. 
Here one may note a classification of such surfaces by Senovilla \cite{Sen}. Of 
concern are normal vectors $\eta$, with $L_\eta$ denoting the Lie derivative 
along $\eta$. This induces a normal 1-form Lie derivative $L$ defined by the 
action 
\begin{equation}
L(\eta)=L_\eta.
\end{equation}
Note that while the normal space need not be integrable, there is a normal 
vector space at each point, and therefore a 1-form space defined by their 
action on such vectors. The {\em expansion 1-form} $\theta$ is defined by 
\begin{equation}
\star\theta=L(\star1)
\end{equation}
where $\star$ is the Hodge operator induced on $S$ by the space-time metric 
$g$, i.e.\ $\star1$ is the area form and $\theta$ its logarithmic normal 
derivative. The {\em expansion vector} (a.k.a.\ mean-curvature vector) is 
\begin{equation}
H=g^{-1}(\theta).
\end{equation}
There is also a Hodge operator $*$ in the normal space, e.g.\ $*\theta$ is the 
{\em dual expansion 1-form}. This induces a duality operation on normal vectors 
by 
\begin{equation}
g(\eta_*)=*g(\eta)
\end{equation}
or equivalently
\begin{equation}
g(\eta_*,\eta)=0,\quad g(\eta_*,\eta_*)=-g(\eta,\eta).
\end{equation}
In particular, there is the {\em dual expansion vector} $H_*$. In spherical 
symmetry, one has $H_*=2k/r$ and $H=2k_*/r$, where $r$ is the area radius and 
$k$ is the Kodama vector \cite{ts1,Kod,sph,1st}. 

A surface is {\em trapped} if $H$ is temporal, or equivalently if $H_*$ is 
spatial. Assuming a time-orientable space-time, the surface is {\em future} 
(respectively {\em past}) trapped if $H$ is future (respectively past) 
temporal.

Now 
\begin{equation}
H_*\cdot\theta=0
\end{equation}
which expresses that any surface is extremal in the $H_*$ direction. In the 
special case $H=0$, the surface is extremal in any normal direction, but 
otherwise $H_*$ gives the unique such normal direction. Thus a trapped surface 
is equivalently defined as a surface which is extremal in a unique spatial 
normal direction. Then it is natural to ask whether the surface is not merely 
extremal but minimal.

Definition 1. A (strictly) {\em minimal} trapped surface is a trapped surface 
for which, for some variation,
\begin{equation}\label{minimal}
Q=H_*^aH_*^b\nabla_b\theta_a>0
\end{equation}
where $\nabla$ is the covariant derivative operator of $g$. Note that 
minimality itself requires only a non-strict inequality, but the strict sign 
will turn out to be convenient. In expressions with indices, there is no need 
to distinguish $\theta$ from $H$, or $*\theta$ from $H_*$, but the distinction 
will be maintained here. 

To avoid cumbersome calculations with connection components, it is useful to 
rewrite the minimality condition as follows. First note that 
\begin{equation}\label{min2}
Q=-H^aH_*^b\nabla_b(*\theta)_a
\end{equation}
due to orthogonality of $H_*$ and $H$. For any 1-form $\alpha$ one has
\begin{eqnarray}
\alpha^b\nabla_b\alpha_a&=&\alpha^b\nabla_a\alpha_b+2\alpha^b\nabla_{[b}\alpha_{a]}\nonumber\\
&=&\textstyle{\frac12}(dg^{-1}(\alpha,\alpha))_a+\alpha^b(d\alpha)_{ba}\label{id}
\end{eqnarray}
where $d$ is the exterior derivative, so that one can do comparatively 
straightforward calculations using exterior calculus. Since the concern here is 
with normal vectors and 1-forms, one can write 
\begin{equation}\label{Q0}
2Q=-H^a(dg^{-1}(\theta,\theta))_a-2H^aH_*^b(d{*}\theta)_{ba}
\end{equation}
where $d$ is henceforth the normal exterior derivative.

\section{Outer trapped surfaces}
In spherical symmetry, outer trapped spheres were defined by $\kappa>0$, where 
$\kappa$ is surface gravity \cite{ts1,1st,ine}. The relevant object in general 
turns out to be 
\begin{equation}
K={*}d{*}\theta+\textstyle{\frac12}g^{-1}(\theta,\theta)
\end{equation}
where ${*}d{*}$ is the normal codifferential or divergence, fixing the sign 
convention. For want of a better term, $K$ will be called the {\em surface 
curvature}. In spherical symmetry, one finds $K=4\kappa/r$. This object has 
appeared before: a previous definition \cite{MH} of quasi-local surface gravity 
was 
\begin{equation}\label{kappa}
\kappa=\frac1{16\pi R}\oint_S\star K
\end{equation}
where $R=\sqrt{A/4\pi}$ is the area radius, i.e.\ area is
\begin{equation}
A=\oint_S\star1=4\pi R^2.
\end{equation}
This $\kappa$ enters a quasi-local first law for trapping horizons \cite{MH} 
involving the Hawking mass \cite{Haw2}. 

Definition 2. An {\em outer} trapped surface is a trapped surface for which, 
for some variation, 
\begin{equation}\label{outer}
K>0.
\end{equation}

Lemma 1. Assuming the Einstein equation with units $G=1$, 
\begin{equation}\label{Q}
2Q=-g^{-1}(\theta,\theta)K-16\pi H\cdot\Psi
\end{equation}
where
\begin{equation}
\Psi=(T+\Theta)\cdot H+w\theta
\end{equation}
where $T$ is the energy tensor,
\begin{equation}
w=-\textstyle{\frac12}\hbox{tr}\,(T+\Theta)
\end{equation}
is an energy density, where the trace is in the normal space, and $\Theta$ is 
an effective energy tensor for gravitational radiation which has appeared in 
various contexts \cite{bhd6,cyl,gwbh,gwe,bhd2,bhd3,bhd5,gr,amf,BHMS,pla}. 

Proof. It is convenient to use a dual-null formalism, describing null 
hypersurfaces generated from the surface in the null normal directions. This 
will not be described in detail here, instead referring to \cite{dne} and, in 
almost the same notation as here, \cite{bhd3}. In terms of null coordinates 
$x^\pm$ and the corresponding null normal vectors $l_\pm$, and with the 
shorthand notation $L_\pm=L_{l_\pm}$, one has 
\begin{eqnarray}
e^{-2\varphi}&=&-g^{-1}(dx^+,dx^-)\\
l_\pm&=&-e^{2\varphi}g^{-1}(dx^\mp)\\
\theta&=&\theta_+dx^++\theta_-dx^-\\
{*}\theta&=&-\theta_+dx^++\theta_-dx^-\\
H&=&-e^{-2\varphi}(\theta_+L_-+\theta_-L_+)\\
H_*&=&e^{-2\varphi}(\theta_+L_--\theta_-L_+).
\end{eqnarray}
Then
\begin{eqnarray}
g^{-1}(\theta,\theta)&=&-2e^{-2\varphi}\theta_+\theta_-\\
d{*}\theta&=&(L_+\theta_-+L_-\theta_+)dx^+\wedge dx^-\\
{*}d{*}\theta&=&-e^{-2\varphi}(L_+\theta_-+L_-\theta_+)\\
K&=&-e^{-2\varphi}(L_+\theta_-+L_-\theta_++\theta_+\theta_-).
\end{eqnarray}
Then (\ref{Q0}) expands as
\begin{eqnarray}
Q&=&e^{-2\varphi}(\theta_-L_++\theta_+L_-)(e^{-2\varphi}\theta_+\theta_-)
-2e^{-4\varphi}\theta_+\theta_-(L_+\theta_-+L_-\theta_+)\nonumber\\
&=&e^{-2\varphi}\left(\theta_-^2L_+(e^{-2\varphi}\theta_+)+\theta_+^2L_-(e^{-2\varphi}\theta_-)\right)
-e^{-4\varphi}\theta_+\theta_-(L_+\theta_-+L_-\theta_+)\nonumber\\
&=&e^{-2\varphi}\left(\theta_-^2\left(L_+(e^{-2\varphi}\theta_+)+\textstyle{\frac12}e^{-2\varphi}\theta_+^2\right)
+\theta_+^2\left(L_-(e^{-2\varphi}\theta_-)+\textstyle{\frac12}e^{-2\varphi}\theta_-^2\right)\right)
-e^{-4\varphi}\theta_+\theta_-(L_+\theta_-+L_-\theta_++\theta_+\theta_-)\qquad{}
\end{eqnarray}
where the last step has added and subtracted terms in 
$e^{-4\varphi}\theta_+^2\theta_-^2$. The reason for this comes from the 
null-null components of the Einstein equation, which can be written as 
\cite{dne,bhd3} 
\begin{equation}\label{Ein}
e^{2\varphi}L_\pm(e^{-2\varphi}\theta_\pm)+\textstyle{\frac12}\theta_\pm^2
=-8\pi T_{\pm\pm}-\textstyle{\frac14}||\sigma_\pm||^2
\end{equation}
where $\sigma_\pm$ are the null shears and $||\sigma_\pm||^2$ their norms with 
respect to the induced metric on the surface. Now the effective energy tensor 
has null-null components \cite{cyl,gwbh,gwe,bhd2,bhd3,bhd5,gr,amf,BHMS,pla} 
\begin{equation}
\Theta_{\pm\pm}=||\sigma_\pm||^2/32\pi
\end{equation}
and $\Theta_{\pm\mp}$ will cancel out. Thus
\begin{equation}
Q=-8\pi e^{-4\varphi}\left(\theta_-^2(T_{++}+\Theta_{++})+\theta_+^2(T_{--}+\Theta_{--})\right)
-e^{-4\varphi}\theta_+\theta_-(L_+\theta_-+L_-\theta_++\theta_+\theta_-).
\end{equation}
Finally one has
\begin{eqnarray}
w&=&e^{-2\varphi}(T_{+-}+\Theta_{(+-)})\\
\Psi&=&-e^{-2\varphi}\left(\theta_-(T_{++}+\Theta_{++})dx^++\theta_+(T_{--}+\Theta_{--})dx^-\right)\label{Psi}
\end{eqnarray}
which implies
\begin{equation}
H\cdot\Psi=e^{-4\varphi}\left(\theta_-^2(T_{++}+\Theta_{++})+\theta_+^2(T_{--}+\Theta_{--})\right).
\end{equation}
Then (\ref{Q}) follows by collecting terms.

Proposition 1. {\em NEC and minimal trapped implies outer trapped.}

Proof. NEC $\Rightarrow H\cdot\Psi\ge0$, as is most easily seen from 
$T_{\pm\pm}\ge0$, $\Theta_{\pm\pm}\ge0$ and the above equation. For a trapped 
surface, $g^{-1}(\theta,\theta)<0$, then inspect signs in (\ref{minimal}), 
(\ref{outer}) and (\ref{Q}). 

In spherical symmetry, one finds $Q=8k^ak^b\nabla_bk_a/r^3$, 
$g^{-1}(\theta,\theta)=4(1-2m/r)/r^2$ and $H\cdot\Psi=4k_*\cdot\psi/r^2$ in 
terms of the mass function $m$ and energy flux $\psi$ 
\cite{ts1,sph,1st,ine,MS}. 

\section{Increasingly trapped surfaces}
Noting that $g^{-1}(*\theta,*\theta)=-g^{-1}(\theta,\theta)$ vanishes for 
marginal surfaces and is positive for trapped surfaces, it can be taken as a 
measure of how trapped a surface is. The idea then is to ask whether it is 
increasing to the future (respectively past) for a future (respectively past) 
trapped surface. 

Definition 3. An {\em increasingly} trapped surface is a trapped surface for 
which, for some variation,  
\begin{equation}\label{increase}
H\cdot dg^{-1}(\theta,\theta)<0.
\end{equation}

Lemma 2. Assuming the Einstein equation, 
\begin{equation}\label{inc}
-H\cdot dg^{-1}(\theta,\theta)=
16\pi H\cdot\Psi-g^{-1}(\theta,\theta)(K-g^{-1}(\theta,\theta)).
\end{equation}

The proof is a special case of the one given in the next section.

Proposition 2. {\em NEC and outer trapped implies increasingly trapped.}

Proof. For a trapped surface, $g^{-1}(\theta,\theta)<0$, and NEC $\Rightarrow 
H\cdot\Psi\ge0$ as before, then inspect signs in (\ref{outer}), (\ref{inc}).

This motivates the following. 

Definition 4. An {\em anyhow increasingly} trapped surface is a future 
(respectively past) trapped surface for which, for all variations along a 
future (respectively past) causal normal vector $\zeta$, 
\begin{equation}
\zeta\cdot dg^{-1}(\theta,\theta)<0.
\end{equation}

Definition 5. A {\em somehow increasingly} trapped surface is a future 
(respectively past) trapped surface for which, for some variation along a 
future (respectively past) causal normal vector $\zeta$, 
\begin{equation}
\zeta\cdot dg^{-1}(\theta,\theta)<0.
\end{equation}

Clearly anyhow increasingly trapped implies increasingly trapped, which implies 
somehow increasingly trapped.

\section{Doubly outer trapped surfaces}
In spherical symmetry, outer trapped implies anyhow increasingly trapped 
\cite{ts1}, but this does not hold in general. Instead, a stricter version of 
outer trapped has this property, as follows. Introduce two more surface 
curvatures: 
\begin{equation}
K_{(\pm)}={*}d{*}\theta\mp{*}d\theta+\textstyle{\frac12}g^{-1}(\theta,\theta)
\end{equation}
where $(\pm)$ indicates a label rather than an index. Then 
$2K=K_{(+)}+K_{(-)}$. 

Definition 6. A {\em doubly outer} trapped surface is a trapped surface for 
which, for all variations, 
\begin{equation}\label{double}
K_{(+)}>0,\quad K_{(-)}>0.
\end{equation}

Clearly doubly outer trapped implies outer trapped. 

Lemma 3. Assuming the Einstein equation, 
\begin{equation}\label{inc2}
-\zeta\cdot dg^{-1}(\theta,\theta)=
16\pi\zeta\cdot\Psi+g^{-1}(\theta,\theta)\zeta\cdot\theta
-\textstyle{\frac12}\zeta\cdot(\theta-{*}\theta)K_{(-)}
-\textstyle{\frac12}\zeta\cdot(\theta+{*}\theta)K_{(+)}.
\end{equation}

Proof. First
\begin{eqnarray}
d\theta&=&(L_+\theta_--L_-\theta_+)dx^+\wedge dx^-\\
{*}d\theta&=&-e^{-2\varphi}(L_+\theta_--L_-\theta_+)\\
K_{(\pm)}&=&-e^{-2\varphi}(2L_\mp\theta_\pm+\theta_+\theta_-).
\end{eqnarray}
Then
\begin{eqnarray}
-\textstyle{\frac12}\zeta\cdot dg^{-1}(\theta,\theta)&=&
(\zeta^+L_++\zeta^-L_-)(e^{-2\varphi}\theta_+\theta_-)\nonumber\\
&=&\zeta^+\theta_-L_+(e^{-2\varphi}\theta_+)+\zeta^-\theta_+L_-(e^{-2\varphi}\theta_-)
+e^{-2\varphi}(\zeta^+\theta_+L_+\theta_-+\zeta^-\theta_-L_-\theta_+)\nonumber\\
&=&\zeta^+\theta_-\left(L_+(e^{-2\varphi}\theta_+)+\textstyle{\frac12}e^{-2\varphi}\theta_+^2\right)
+\zeta^-\theta_+\left(L_-(e^{-2\varphi}\theta_-)+\textstyle{\frac12}e^{-2\varphi}\theta_-^2\right)\nonumber\\
&&\quad+e^{-2\varphi}\left(\zeta^+\theta_+(L_+\theta_--\textstyle{\frac12}\theta_+\theta_-)
+\zeta^-\theta_-(L_-\theta_+-\textstyle{\frac12}\theta_+\theta_-)\right)\label{inc3}
\end{eqnarray}
where terms have been added and subtracted in the last step in order to use the 
null-null Einstein equations (\ref{Ein}) again. Then 
\begin{eqnarray}
-\textstyle{\frac12}\zeta\cdot dg^{-1}(\theta,\theta)&=&
-8\pi e^{-2\varphi}\left(\zeta^+\theta_-(T_{++}+\Theta_{++})
+\zeta^-\theta_+(T_{--}+\Theta_{--})\right)
-e^{-2\varphi}\theta_+\theta_-(\zeta^+\theta_++\zeta^-\theta_-)\nonumber\\
&&\quad+e^{-2\varphi}\left(\zeta^+\theta_+(L_+\theta_-+\textstyle{\frac12}\theta_+\theta_-)
+\zeta^-\theta_-(L_-\theta_++\textstyle{\frac12}\theta_+\theta_-)\right)
\end{eqnarray}
where more terms have been added and subtracted. The terms in 
$\zeta^\pm\theta_\pm$ can be written invariantly using
\begin{equation}
\theta\mp{*}\theta=2\theta_\pm dx^\pm.
\end{equation}
Finally (\ref{Psi}) gives 
\begin{equation}
\zeta\cdot\Psi=-e^{-2\varphi}\left(\zeta^+\theta_-(T_{++}+\Theta_{++})
+\zeta^-\theta_+(T_{--}+\Theta_{--})\right)
\end{equation}
and (\ref{inc2}) follows.

Proposition 3. {\em NEC and doubly outer trapped implies anyhow increasingly 
trapped.} 

Proof. For a trapped surface, $g^{-1}(\theta,\theta)<0$, while for $\zeta$ in 
the appropriate causal quadrant, $\zeta\cdot\theta<0$, 
$\zeta^\pm\theta_\pm\le0$ and $\zeta^\pm\theta_\mp\le0$, so 
$\zeta\cdot(\theta\pm{*}\theta)\le0$, then NEC $\Rightarrow 
\zeta\cdot\Psi\ge0$, as is most easily seen from the above equation. Then 
({\ref{double}) in (\ref{inc2}) implies $\zeta\cdot dg^{-1}(\theta,\theta)<0$ 
for all such $\zeta$. 

Taking $\zeta=H$ proves Lemma 2. The proof also makes clear that outer trapped 
generally does not imply anyhow increasingly trapped. 

\vfill\eject
\section{Involute trapped surfaces}
Given the above result, doubly outer trapped seems to be a natural condition, 
so one might ask whether it is implied by minimal trapped. The answer is 
negative, but if the latter is refined further, such a result can be obtained. 
The form (\ref{min2}) motivates the following. 

Definition 7. An {\em involute} trapped surface is a future (respectively past) 
trapped surface for which, for all variations along a future (respectively 
past) causal normal vector $\zeta$, 
\begin{equation}\label{involute}
Q=-\zeta^aH_*^b\nabla_b(*\theta)_a>0.
\end{equation}

Clearly involute trapped implies minimal trapped. Involute means curved or 
curled inwards, as of a leaf, in this case referring to the curvature of the 
surface in the space-time. The Latin root {\em involvere} means to enwrap or 
conceal, as also appropriate in the context of black holes. 

Lemma 4. Assuming the Einstein equation, 
\begin{equation}\label{inv}
-2Q=16\pi\zeta\cdot\Psi
+\textstyle{\frac12}\zeta\cdot(\theta-{*}\theta)K_{(+)}
+\textstyle{\frac12}\zeta\cdot(\theta+{*}\theta)K_{(-)}.
\end{equation}

Proof. Using the identity (\ref{id}) again,
\begin{equation}
-Q=\textstyle{\frac12}\zeta^a(dg^{-1}(\theta,\theta))_a+\zeta^aH_*^b(d{*}\theta)_{ba}.
\end{equation}
The second term expands as
\begin{equation}
\zeta^aH_*^b(d{*}\theta)_{ba}
=-e^{-2\varphi}(\zeta^+\theta_++\zeta^-\theta_-)(L_+\theta_-+L_-\theta_+)
\end{equation}
while the first term has been calculated above (\ref{inc3}). Adding, one finds 
that two contributions from the second term cancel, leaving 
\begin{eqnarray}
-Q&=&-8\pi e^{-2\varphi}\left(\zeta^+\theta_-(T_{++}+\Theta_{++})
+\zeta^-\theta_+(T_{--}+\Theta_{--})\right)\nonumber\\
&&\quad-e^{-2\varphi}\left(\zeta^+\theta_+(L_-\theta_++\textstyle{\frac12}\theta_+\theta_-)
+\zeta^-\theta_-(L_+\theta_-+\textstyle{\frac12}\theta_+\theta_-)\right).
\end{eqnarray}
Collecting terms yields (\ref{inv}).

Proposition 4. {\em NEC and involute trapped implies doubly outer trapped.} 

Proof. As before, $\zeta^\pm\theta_\pm\le0$ and NEC $\Rightarrow 
\zeta\cdot\Psi\ge0$. Considering the cases $\zeta=l_\pm$, both $K_{(\pm)}$ must 
be positive in (\ref{inv}). 

The proof also makes clear that minimal trapped generally does not imply doubly 
outer trapped. 

Definition 8. An {\em somehow involute} trapped surface is a future 
(respectively past) trapped surface for which, for some variation along a 
future (respectively past) causal normal vector $\zeta$, 
\begin{equation}\label{involute}
Q=-\zeta^aH_*^b\nabla_b(*\theta)_a>0.
\end{equation}

Clearly minimal trapped implies somehow involute trapped. One might ask whether 
somehow involute trapped implies somehow increasingly trapped. The answer is 
negative, except for a special case. 

Proposition 5. {\em NEC and somehow involute trapped implies somehow 
increasingly trapped if $K_{(+)}=K_{(-)}$.} 

Proof. From (\ref{inv}) one gets 
\begin{equation}
\zeta^+\theta_+K_{(+)}+\zeta^-\theta_-K_{(-)}<0.
\end{equation}
However, to get increasingly trapped in (\ref{inc2}) instead generally requires
\begin{equation}
\zeta^+\theta_+K_{(-)}+\zeta^-\theta_-K_{(+)}<0.
\end{equation}
Thus it works if $K_{(+)}=K_{(-)}$. 

\vfill\eject
\section{Remarks}
Before concluding, one issue should be addressed. Concerning marginal surfaces, 
say with $\theta_+=0$, the author has previously proposed defining an outer 
trapping horizon by $L_-\theta_+<0$, as this naturally expresses that the 
vanishing expansion decreases in an inward direction, and implies some basic 
properties expected of black holes, such as spherical topology, achronality or 
one-way traversability, and a second law, assuming either the dominant energy 
condition or just NEC \cite{bhd,bhd6,bhd2,bhd3,bhd5}. This is equivalent to 
$K_{(+)}>0$. Conversely, for a trapping horizon with $\theta_-=0$, one would 
use $K_{(-)}>0$. On the other hand, for a trapped surface, there is no reason 
to break the symmetry between $K_{(\pm)}$, so the natural conditions are either 
$K>0$ or both $K_{(\pm)}>0$. Thus if one wants a unified treatment of both 
trapped and marginal surfaces, and surfaces which are partly trapped and partly 
marginal, in the context of black holes, it suggests the doubly outer condition 
for all. Thus if the motivating conjecture turns out to be true, that is, the 
boundary of some suitably refined trapped region is a trapping horizon, one 
might further conjecture that it is a doubly outer trapping horizon, i.e.\ both 
$K_{(\pm)}>0$. 

In summary, the hierarchy of trapped surfaces is illustrated as follows:
\begin{eqnarray}
&&\hbox{involute}\Rightarrow\hbox{doubly outer}\Rightarrow\hbox{anyhow increasingly}\nonumber\\
&&\Downarrow\qquad\hbox{NEC}\qquad\Downarrow\qquad\hbox{NEC}\qquad\Downarrow\nonumber\\
&&\hbox{minimal}\quad\Rightarrow\quad\hbox{outer}\quad\Rightarrow\quad\hbox{increasingly}\nonumber\\
&&\Downarrow\qquad\quad\hbox{NEC}\qquad\qquad\hbox{NEC}\qquad\quad\Downarrow\nonumber\\
&&\hbox{somehow involute}\quad\Rightarrow\quad\hbox{somehow increasingly}\nonumber\\
&&\qquad\qquad\qquad\hbox{NEC, $K_{(+)}=K_{(-)}$} 
\end{eqnarray}
where the vertical implications are straightforward, while the horizontal 
implications require NEC, and in the last case, the symmetry where the 
curvatures $K_{(\pm)}$ are equal, or equivalently where outer trapped implies 
doubly outer trapped, in which case the hierarchy collapses to a strict 
hierarchy as in spherical symmetry \cite{ts1}. Otherwise, to borrow language of 
warp and weft from weaving, the threads are respectively geometrical warp and 
physical weft. 

\medskip
Thanks to Ingemar Bengtsson and Jos\'e Senovilla for discussions. Research 
supported by the National Natural Science Foundation of China under grants 
10375081, 10473007 and 10771140, by Shanghai Municipal Education Commission 
under grant 06DZ111, and by Shanghai Normal University under grant PL609. 

\vfill\eject

\end{document}